# Unraveling the Global Teleconnections of Indian Summer Monsoon Clouds: Expedition from CMIP5 to CMIP6

Ushnanshu Dutta[1,2], Anupam Hazra[1*], Hemantkumar S. Chaudhari[1], Subodh Kumar Saha[1], Samir Pokhrel[1], and Utkarsh Verma[1]

[1]Indian Institute of Tropical Meteorology, Ministry of Earth Sciences, India.

[2]Department of Atmospheric and Space Sciences, Savitribai Phule Pune University, Pune, India.

Corresponding author: Anupam Hazra (hazra@tropmet.res.in)

**Key Points:**

- The total cloud fraction (TCF) over the Indian summer monsoon(ISM) region is strongly associated with global predictors.

- The seasonal mean bias of TCF and rainfall over the Asian Summer Monsoon Region have seen improvement from CMIP5-MME to CMIP6-MME.

- Improvement from CMIP5 to CMIP6 models may be attributed to the better depiction of the observed global teleconnections.

**Abstract**

We have analyzed the teleconnection of total cloud fraction (TCF) with global sea surface temperature (SST) in multi-model ensembles (MME) of the fifth and sixth Coupled Model Intercomparison Projects (CMIP5 and CMIP6). CMIP6-MME has a more robust and realistic teleconnection (TCF and global SST) pattern over the extra-tropics (R ~0.43) and North Atlantic (R ~0.39) region, which in turn resulted in improvement of rainfall bias over the Asian summer monsoon (ASM) region. CMIP6-MME can better reproduce the mean TCF and have reduced dry (wet) rainfall bias on land (ocean) over the ASM region. CMIP6-MME has improved the biases of seasonal mean rainfall, TCF, and outgoing longwave radiation (OLR) over the Indian Summer Monsoon (ISM) region by ~40%, ~45%, and ~31%, respectively, than CMIP5-MME and demonstrates better spatial correlation with observation/reanalysis. Results establish the credibility of the CMIP6 models and provide a scientific basis for improving the seasonal prediction of ISM.

**Plain Language Summary**

Skillful seasonal prediction of monsoon has great socio-economic value. Indian summer monsoon rainfall (ISMR) prediction remained a challenging problem with the sub-critical skills of the dynamical models. This arises from the limited understanding of the interaction among clouds and convection. We find significant improvement in TCF climatology in the CMIP6 model and a better teleconnection pattern with global SST. The large-scale convection and rainfall in CMIP5, CMIP6 models, and their ensembles over the ASM and ISM region are also evaluated. The improved physical representation of clouds in models has led to a better mean and teleconnection relative to the previous generation of models. Therefore, to provide a more realistic monsoon simulation, it is crucial to have new generation models, which most likely represent the improved relationship of clouds and SST than their predecessors.

## 1. Introduction

During the few decades, progress in climate modeling has provided new insight regarding ASM. The interannual and intraseasonal variability of ASM in general and Indian summer monsoon (ISM) rainfall (ISMR) in particular controls the livelihood and health of over two billion people and regulates the country's economy through food production (Gadgil, 2003, Gadgil and Rupa Kumar, 2006, Gupta et al., 2019, Krishna Kumar et al., 2004). Hence, predicting this variability well in advance can help farmers in crop management and the government prepare for natural calamities (e.g., flood and drought).

The continuous efforts on development in model resolution and physics have shown advancement in ASM rainfall simulation in the CMIP6 (Eyring et al., 2016) models relative to the CMIP5 (Taylor et al., 2012) and CMIP Phase 3 (CMIP3, Meehl et al., 2007) models. Sperber et al. (2013) and Seo et al. (2013) have compared CMIP5 and CMIP3 model ensembles and pointed out that CMIP5 displayed better skills over ASM in many aspects of monsoon diagnostics. Gusain et al. (2020), in a preliminary study, compared a few CMIP5 and CMIP6 models regarding their spatiotemporal variation over the ISM region. Recent studies have compared CMIP5 and CMIP6 in simulating mean rainfall over east ASM (Xin et al., 2020) and east Africa (Ayugi et al., 2021) region. Zhu and Yang (2021) also assessed the CMIP5 and CMIP6 in terms of simulated interdecadal and interannual variation of global precipitation. Vignesh et al. (2020) found the lacuna of similar assessment studies regarding cloud distribution globally. They have compared CMIP5 and CMIP6 models based on seasonal and regional variations of cloud fractions. Several researchers (Chen et al., 2020; Tokarska et al. 2020; Zelinka et al. 2020; Nie et al. 2020) also carried out similar assessment studies for different variables (e.g., temperature, circulation, etc.). These studies generally highlight the improvement of mean structure in CMIP6 participating models from CMIP5 models with certain exceptions in different regions across the globe.

However, despite the improvement in the simulation of the mean structure of ISMR, prediction-skill is still below the potential limit for the models (Kumar et al., 1999; Rajeevan et al., 2012). Tiwari et al. (2014) found that these low skill scores may originate from the sparse representation of observed teleconnection of ISMR with Pacific SST, i.e., El-Nino and Southern Oscillation (ENSO), revealed in numerous studies (Charney and Shukla, 1981; Krishnamurthy and Goswami, 2000; Ropelewski and Halpert, 1989; Sikka 1980; Mooley and Parthasarathy 1984; Pradhan et al., 2016; Goswami and Xavier 2005; Goswami and Jayavelu 2001; Pokhrel et al., 2012; Dwivedi et al., 2015; Saha et al., 2019). Roy et al. (2019) and Mahendra et al. (2021) tested the fidelity of this ENSO-ISMR relationship from CMIP5 and CMIP6 models, respectively but with a different approach. Contrastingly, the climatic relationship or teleconnection of ISMR with other potential sources of predictability such as North Atlantic Oscillations (NAO), Atlantic Multidecadal Oscillations (AMO), and Extra Tropics (ET) seems to gain less attention comparatively. (Burns et al., 2003; Srivastava et al., 2002; Chang et al., 2001; Chattopadhyay et al., 2015; Sankar et al., 2016; Borah et al., 2020). Several studies have discussed the structure and variability of these modes (viz. NAO, AMO, ENSO, etc.) from CMIP3 or CMIP5 models (Medhaug and Furevik, 2011, Ting et al., 2011; Zhang and Wang 2013; Wang et al., 2017, etc.) and Fasullo (2020) have assessed their performances with CMIP6 models. However, the teleconnection studies of ISMR with these sources (e.g., AMO) from CMIP models are found in limited studies (Luo et al., 2018; Joshi and Ha, 2019).

On the other hand, all these kinds of teleconnection studies regarding ISM clouds are not focused in detail from both observational and modeling aspects, despite clouds play a seminal role in governing rainfall variability (Chaudhari et al., 2016) through modulation of heating (Hazra et al., 2017a, b; Hong et al., 2016; Baker, 1997) and induced circulation (Kumar et al., 2014). The individual cloud complexes are also revealed by satellite-derived cloud data (Wang and Rui, 1990). Nakazawa (1988) showed that the intraseasonal variability (ISV) is related to the large-scale cloud complexes. Therefore, earlier studies (Stephens et al., 2002; Hazra et al., 2015, 2016, 2017a, b, 2020; Waliser et al., 2009) highlighted that clouds regulate radiative energy and maintain water cycle balances. Bony et al. (2015) reported that clouds are essential for climate sensitivity studies in climate models. However, the misrepresentation of clouds, precipitation, and circulation has been continued for many new-generation models (Morrison et al., 2020). De et al. (2019) emphasized that the interaction between cloud and large-scale circulation remains a 'grey area of climate science'. Cloud microphysical processes also play a significant role in modulating the ISV of ISM (Bony et al., 2015; Kumar et al., 2017; Hazra et al., 2017a, 2020; Dutta et al., 2020, 2021). Therefore, the role of clouds in general and cloud condensates like cloud ice, in particular, is essential for the simulation of subseasonal disturbances by AOGCMs (Atmosphere-Ocean General Circulation Models) with higher fidelity (Dutta et al., 2020; Hazra et al. 2017a, b, 2020). This is also a critical requirement for the seasonal prediction of south Asian monsoon rainfall in particular and tropics in general.

Therefore, it is essential to explore whether the interannual variability of ISM clouds is also teleconnected with the slowly varying predictable component like SST around the globe. Assessment of the simulated relationship from the recent two generations of CMIP will also provide new insight for the scientific community in model development. Recently, Kim et al. (2020) evaluated the performance of CMIP5 and CMIP6 models based on the observed teleconnection of heatwaves over Korea. This kind of assessment study of CMIP5 and CMIP6 has been overlooked for the teleconnection of ISM clouds (and rainfall) with global predictors along with its mean structure. Therefore, it is time for this study to evaluate the performance of CMIP6 ensembles compared to CMIP5 in simulating the ASM in general and ISM clouds in particular. In this study, we have discussed the following:

i) Assessment of mean structure of ASM and ISM with respect to rainfall, clouds, and convection from ensembles of CMIP5 and CMIP6. We have considered more numbers of models (30 for each) than previous studies (Gusain et al., 2020, Vignesh et al., 2020).

ii) Exploring the vital role of cloud fractions for ASM (ISM in particular) through teleconnection with global predictors (ENSO, NAO, AMO, and ET) and comparative evaluation of this teleconnection from CMIP5 and CMIP6 models and their respective MME.

iii) Similar assessment of simulated teleconnection of ISM rainfall from CMIP5 and CMIP6.

## 2. Data and Methods

### 2.1. Observational and reanalysis datasets:

Monthly data of TCF are obtained from the GCM-Oriented CALIPSO Cloud Product (GOCCP, Chepfer et al., 2010) for the available period during 2006-2017 (12 years). This product is widely used to evaluate 'models' cloudiness (Cesana and Chepfer, 2012; 2013). Monthly data of TCF from the recently released fifth generation of the European Centre for Medium-Range Weather Forecasts (ECMWF) reanalysis, ERA5 (Hersbach et al. 2020) are considered for long data sets (30 years), which is not available from satellite observation. It uses ECMWF's Earth System model IFS, cycle 41r2 (Pokhrel et al. 2020). For observed rainfall, monthly data are taken from the Global Precipitation Climatology Project Version 2.3 (GPCP, Adler et al., 2003), and daily data are taken from India Meteorological Department (IMD, Rajeevan et al., 2006). High-resolution quality-controlled monthly outgoing longwave radiation (OLR) at the top of the atmosphere data from geostationary platforms like Kalpana-1(formerly METSAT-1, Mahakur et al., 2013) are considered in this study. We have taken SST data from the Hadley Centre Global Sea Ice and Sea Surface Temperature (HadiSST, Rayner et al., 2003) and OLR from the National Oceanic and Atmospheric Administration (NOAA) (Liebmann and Smith, 1996). Most of the CMIP5-models data are available until 2005, whereas GPCP and ERA5 data are available from 1979. Hence for 30 years long data period, 1979-2008 is chosen from observation (GPCP, IMD) and reanalysis (ERA5, NOAA, HadiSST).

## 2.2. CMIP5 and CMIP6 model datasets

Monthly data of TCF, rainfall, SST, and OLR are obtained from thirty CMIP5 and CMIP6 models for the common period (1976-2005, 30 years) of historical runs for all the model simulations. Multi-model ensemble of CMIP5 (CMIP5-MME) and CMIP6 (CMIP6-MME) are generated from the common period data for the present analysis. We have performed 'regridding' on several CMIP5 and CMIP6' models' SST data to transform from curvilinear to the fixed latitude-longitude grid (1º x 1º) using bilinear-interpolation method (https://climatedataguide.ucar.edu/climate-data-tools-and-analysis/regridding-overview).

To evaluate the spatial teleconnection pattern, JJAS mean of Central India (CI: 72-88ºE, 18-28ºN) averaged rainfall and TCF are correlated with JJAS mean global (at each grid point) SST. The correlations are further quantified over different boxes across the globe, i.e., Niño (160˚E-80˚W, 5˚S-5˚N), NAO (80ºW-30ºE, 35ºN-65ºN, Athanasiadis et al., 2020), AMO (80ºW-0ºE,0-65ºN, https://climatedataguide.ucar.edu/climate-data/atlantic-multi-decadal-oscillation-amo), and ET (100ºE-5ºW, 15ºN-75ºN, Chattopadhyay et al., 2015) to evaluate the teleconnection strength. Pattern correlation (or map to map correlation, Lund, 1963) coefficient (PCC) is also calculated between two variables to assess their spatial consistency quantitatively.

## 3. Results

### 3.1. Mean state of rainfall, convection, and clouds:

The June to September (JJAS) climatology (30 years) of rainfall, OLR, and TCF, from observations/reanalysis, CMIP5-MME, and CMIP6-MME over the ASM region (50-120ºE, 10ºS-40ºN) are presented in Figure 1. The seasonal mean rainfall from GPCP and IMD (only land region) is shown in Figure 1a, d. The maxima of rainfall are seen along the Western Ghats and the Myanmar coast extending to north-east India and the Bay of Bengal (BoB) (Fig. 1a). Central India (CI) and the equatorial eastern Indian Ocean (EEIO) also experience considerable rainfall. The underestimation of rains along the western coast and over BoB by CMIP5-MME (Fig. 1b) is realistically improved in CMIP6-MME (Fig. 1c). Over the ASM region, the CMIP5-MME (5.28 mm/day) overestimates the mean observed rainfall (GPCP: 4.85 mm/day) more than that of CMIP6-MME (5.24 mm/day). The PCC of CMIP6-MME (~0.87) with GPCP over the ASM region is also slightly higher than CMIP5-MME (~0.86).

The climatology of OLR, a proxy of convection (Murakami, 1980a, b; Prasad and Verma 1985), is shown from NOAA (Fig. 1e) and Kalpana satellite product (Fig. 1h) and further compared with CMIP-MMEs. The climatology of OLR from NOAA (Fig. 1e) is spatially well comparable (PCC ~0.98) with Kalpana-1 satellite data (Fig. 1h) over the ASM region. CMIP5-MME (242.1 Watt/m$^2$) and CMIP6-MME (241.1 Watt/m$^2$) both overestimate the mean OLR (NOAA, 234.3 Watt/m$^2$) over the ASM region. The overestimation is reduced in CMIP6-MME (Fig. 1g) and is a little closer to the observed pattern (PCC~0.95) than CMIP5-MME (Fig. 1f, PCC~0.93).

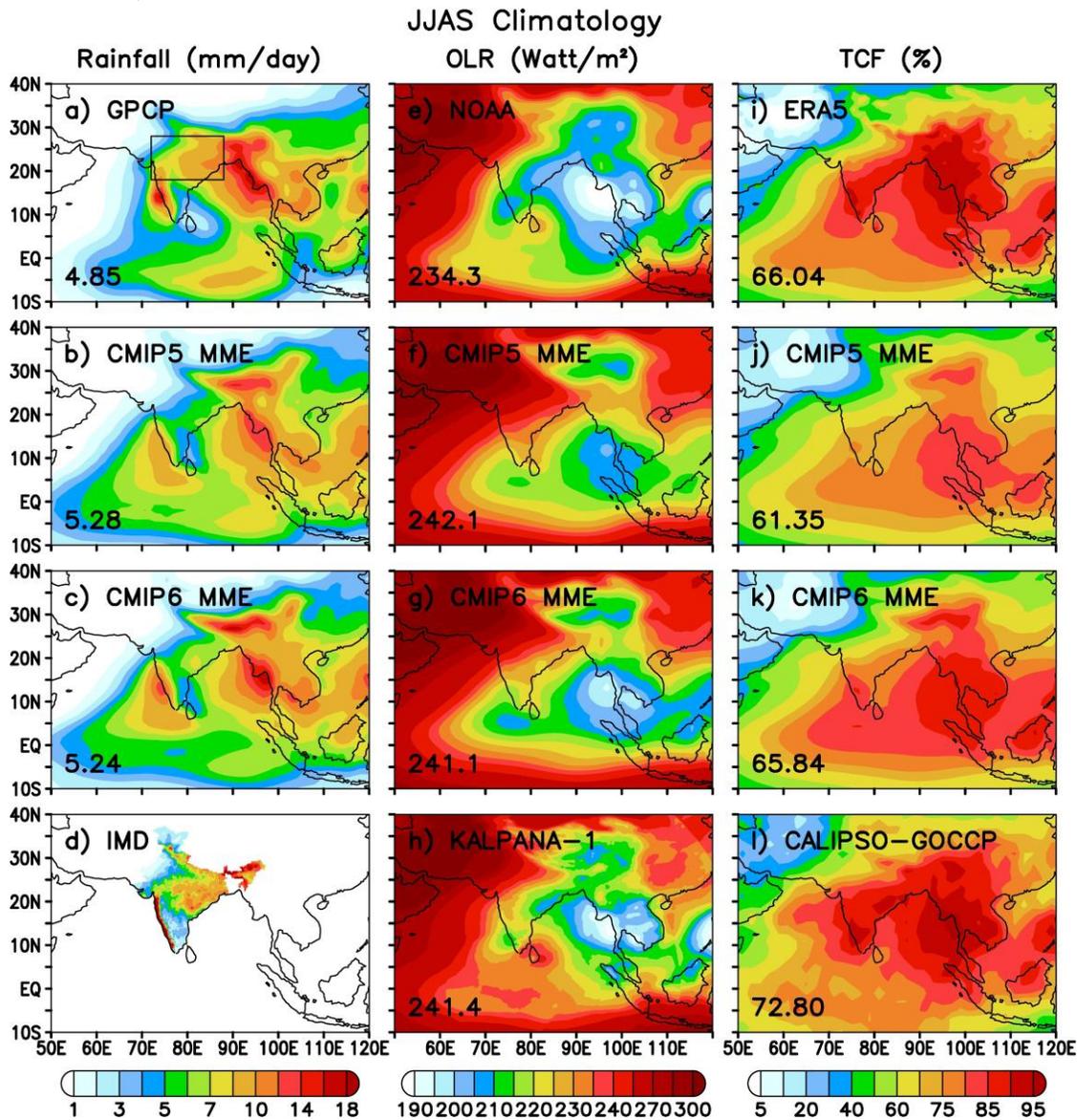

**Figure 1:** *JJAS climatology of rainfall from (a) GPCP, (b) CMIP5-MME, (c) CMIP6-MME, (d) IMD. Outgoing Long-wave Radiation (OLR) from (e) NOAA, (f) CMIP5-MME, (g) CMIP6-MME, (h) KALPANA-1. Total Cloud*

*Fraction (TCF) from (i) ERA5, (j) CMIP5-MME, (k) CMIP6-MME, (l) CALIPSO-GOCCP. Central India box is shown in (a). Basin (ASM region) averaged values of respective variables are written in lower-left corner.*

The spatial distribution of climatological TCF from ERA5 (Fig. 1i) is in good agreement with satellite observation, CALIPSO-GOCCP (Fig. 1l) over the ASM as well as ISM region (PCC ~ 0.96). Hence, we have considered 30 years ERA5 product for evaluating the TCF of two generations model ensembles. The TCF from ERA5 and CALIPSO-GOCCP data (Fig. 1i, l) is spatially well associated (PCC > 0.6) with rainfall over the ASM region. The mean TCF over ASM has been realistically increased in CMIP6-MME (65.84%, Fig. 1k) compared to CMIP5-MME (61.35%, Fig. 1j). However, their spatial distribution over the ASM region is consistent with ERA5 while, PCC is slightly higher in CMIP6-MME (0.97) than CMIP5-MME (0.95). The PCC of these variables (rainfall, OLR, TCF) with observation/reanalysis over ISM region (70-90ºE, 10-30ºN) are also slightly better in CMIP6-MME (0.68, 0.93, 0.96) than CMIP5-MME (0.67, 0.91, 0.94). Hence, we can see significant and realistic improvement of the mean characteristics in CMIP6-MME compared to CMIP5-MME over the ASM region.

**3.2. Teleconnection of ISM rainfall and TCF with Global SST**

Now, we will discuss the relation of ISM rainfall and TCF with global SST from MME of CMIP5 and CMIP6. This will help to get the potential predictability of ISMR. The correlation of CI averaged rainfall (GPCP) with global SST (HadiSST) at each grid point is shown in figure 2a, which is already revealed by many earlier studies (Kripalani and Kulkarni, 1997; Sikka 1980; Kane 1998, 2000; Goswami, 1998 and references therein). The correlation between rainfall and SST from CMIP5-MME (Fig. 2b) and CMIP6-MME (Fig. 2c) are also presented.

To evaluate the role of TCF, we have also calculated the correlation of JJAS mean TCF (averaged over CI) from ERA5 with global SST (HadiSST, Fig. 2d) and compared the same with CMIP5-MME (Fig. 2e) and CMIP6-MME (Fig. 2f). The correlation values are strongly negative over the eastern Pacific region (Fig. 2d). It also shows a 'horseshoe' like significant and positive correlation pattern over the western pacific region. Besides, it resembles a canonical ENSO-like pattern which may imply the modulation of TCF over ISM through the teleconnection with ENSO. A strong and significant positive correlation over the Indian Ocean is also noticed near the Australian coast. A strong correlation with SST over the Atlantic Ocean highlights the importance of the teleconnection of TCF with NAO and AMO (Fig. 2d).
Interestingly, the observed TCF - SST teleconnection (Fig. 2d) is spatially well consistent (PCC~ 0.80) with that of rainfall-SST (Fig. 2a) over the whole globe (0-360, 90˚S-90˚N). This implies that both TCF and rainfall over ISM have similar interactions with global SST. However, the teleconnection strength (i.e., correlation value) is weaker in rainfall-SST over the Pacific region. This may be due to the observed weak relationship between ISMR-ENSO during 1979-1997 (Yang and Huang, 2021; Kumar et al., 1999), and our study period overlaps with that.

The spatial plot of one point correlation between CI rainfall/TCF and SST shows that two generations' models can capture the negative correlation over the eastern and central Pacific as revealed by the observation (Fig. 2a, d). However, the teleconnections in the extra-tropics (Chattopadhyay et al. 2015) and the Atlantic region are better captured by CMIP6-MME (Fig. 2c, f).

It is also noted that the North Atlantic region is essential for non-ENSO drought prediction (Borah et al., 2020). This can be attributed to improved teleconnection between TCF and SST (Fig. 2f) and better agreement with reanalysis. The TCF and global SST teleconnection patterns from individual 30 models of CMIP5 (Fig. S1) and CMIP6 (Fig. S2) are also evaluated. Results also demonstrate the improvement in teleconnection pattern from models participating in CMIP5 to CMIP6.

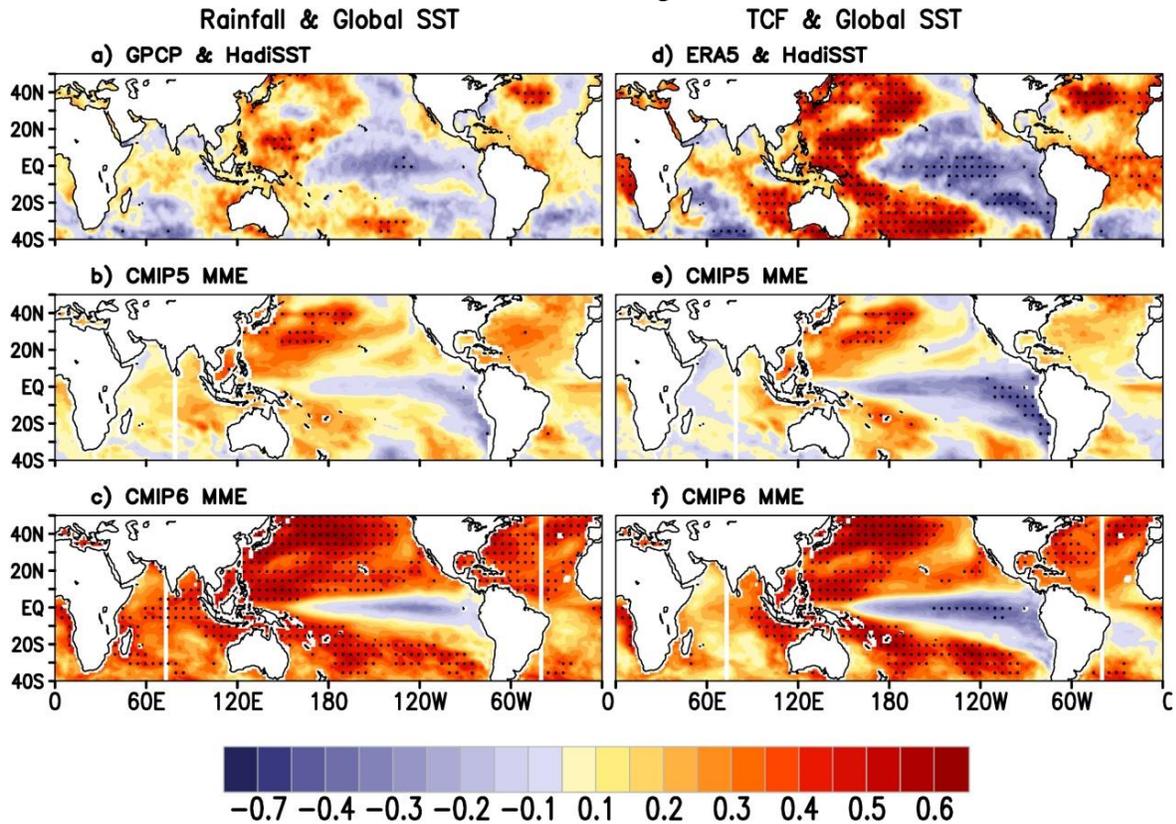

**Figure 2:** *Correlation of central India (CI) averaged rainfall with global SST at each grid point from (a) Observation/reanalysis (GPCP - HadiSST), (b) CMIP5-MME, (c) CMIP6-MME. Similar, correlation between CI averaged total cloud fraction (TCF) and SST from (d) ERA5-HadiSST, (e) CMIP5-MME, (f) CMIP6-MME. Correlation values greater than 95% significance are stippled.*

**3.3. Quantitative evaluation of teleconnection in CMIP5 and CMIP6**

As a quantitative analysis, the correlation coefficient (CC) of CI averaged TCF (and rainfall) with SST of four different boxes across the globe (NINO, NAO, ET, AMO) are presented (Fig. 3). Rainfall from observation and TCF from reanalysis yield negative (positive) CC with SST over Niño (ET, NAO, and AMO) boxes. The opposite correlation of TCF/rainfall with ENSO and NAO suggests that La- Niña (El- Niño) conditions are linked with positive (negative) NAO patterns (Fereday et al., 2020, Goswami et al., 2006). The CC values are more in the TCF-SST case. Overall the CC values are stronger in the CMIP6 participating models. Regarding TCF-SST teleconnection, CC values over ET, AMO and NAO in CMIP6-MME (0.43, 0.36, and 0.39) are closer to reanalysis (0.41, 0.41, and 0.51) than that of CMIP5-MME (0.21, 0.18, and 0.22) (Fig. 3b, d). In the Niño region, both the MMEs obtain lower CC (~ -0.30) than the reanalysis (-0.39). Overall significant progress has been noticed in CMIP6 participating models from CMIP5 in representing the observed teleconnection of TCF with ENSO and other predictors (Fig. 3, S1, and S2).

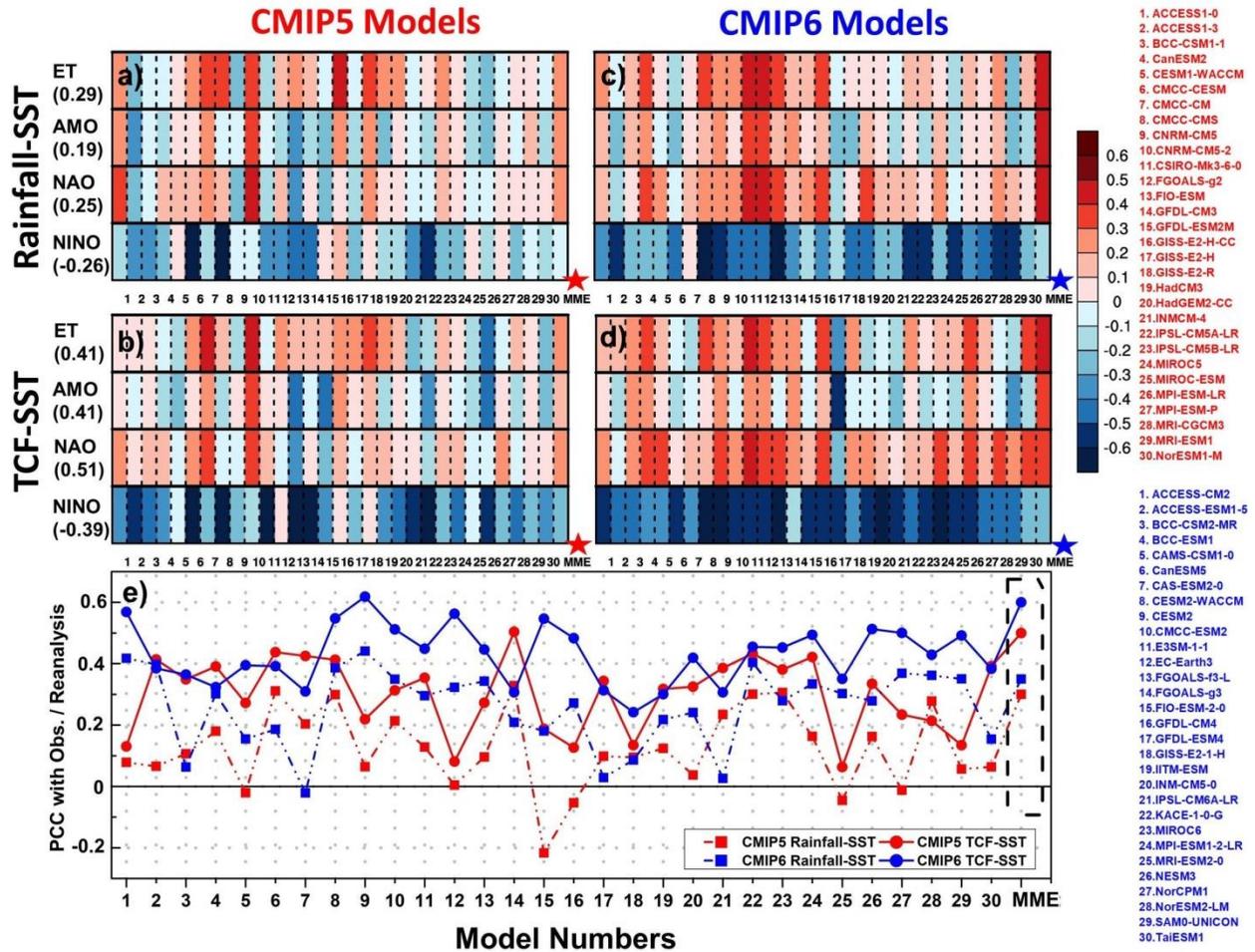

**Figure 3:** *Quantification of teleconnection with different global predictors (NINO, NAO, AMO, and ET) from thirty CMIP5 and CMIP6 models and their MMEs. (a) Rainfall and SST (b) TCF and SST from CMIP5 models. (c) Rainfall and SST (d) TCF and SST from CMIP6 models. The values in the parenthesis are the observed/reanalysis values. (e) Pattern Correlation of all thirty CMIP5, CMIP6 models and their MMEs with observation/ reanalysis across the globe. CMIP5 (CMIP6) model names are written in red (blue) color serially at the rightmost side.*

In general, ENSO and AMO explain a significant fraction of interannual variability (IAV) of ISMR (Borah et al., 2020). The improved correlation of CMIP6 models as compared to CMIP5 along with their respective MME establishes the improved teleconnection. To understand the progress quantitatively, we have calculated the PCC (Fig. 3e) of the teleconnection pattern simulated by each model and CMIP5/CMIP6-MME with observation/reanalysis over the whole globe. It shows that PCC is higher in the CMIP6-MME than CMIP5-MME for both TCF-SST and rainfall-SST teleconnections. PCC is generally higher for the TCF-SST case than the rainfall-SST case for both the generations' models. In both the cases, majority of CMIP6 participating models have higher PCC than that of CMIP5. These results pinpoint the progress of representation of ISM clouds/rainfall and global SST teleconnections, from CMIP5 to CMIP6, encouraging monsoon prediction research in the scientific community.

### 3.4. Improvements in ISM mean characteristics from CMIP5 to CMIP6

Earlier studies (Chaudhari et al., 2013; Saha et al., 2013; Hazra et al., 2015; Jain et al., 2019; Wang et al., 2018) have shown that most of the models simulate significant dry (wet) bias over land (ocean) for ISM region which is identified as a generic problem of models by Rajeevan and Nanjundiah (2009). Can the development of TCF in CMIP6 play a significant role in improving this mean pattern? To explore it, we have presented the scatter plot between seasonal (JJAS) mean rainfall bias and TCF bias (both averaged over CI) for each model and their respective MME (Fig. 4a). The rainfall (TCF) biases are with respect to GPCP (ERA5). The linear relationship between these biases (Fig. 4a) highlights the importance of the proper simulation of TCF to simulate the mean ISMR in models properly. We noticed that few CMIP5 models exhibit less bias in TCF and rainfall, whereas considerable progress is seen in CMIP6 models. The majority of the CMIP6 models have shown improvement in simulating the mean TCF and rainfall (Fig. 4a). The improvement is also seen in the TCF averaged over the ISM region (Fig. 4b). The CMIP6-MME reduces the TCF-bias of CMIP5-MME by approximately 45% over this region. To understand any improvement in capturing the large-scale convection, we compared the mean seasonal OLR simulated by the models with NOAA-OLR data (Fig. 4c) averaged over BoB region (80-100˚E, 10-30˚N), which is the convective heat source for ISM (Joseph and Sijikumar 2004). Results show that OLR in CMIP6-MME (218.85 Watt/m$^2$) is closer to the NOAA (204.7 Watt/m$^2$) as compared to CMIP5-MME (225.31 Watt/m$^2$). The reduction of bias in TCF (~45%) and OLR (~31%) by CMIP6-MME have resulted in a decrease (~40%) of dry bias over the ISM region, too (Fig. 4d). Our results are also in line with recent studies (Vignesh et al., 2020; Gusain et al., 2020). In all of these cases (Figs. 4b-d), more CMIP6 participating models are closer to the observed/reanalysis value. We also noticed that the ensemble spread of CMIP5 models is more than that of CMIP6 for all of these variables. We also investigated the biases of rainfall over land and ocean separately over the ASM region to demonstrate where the improvement took place (Fig. 4e). The CMIP6 participating models and their MME indeed depicts progress in both areas (i.e., land and ocean), and the underestimation (overestimation) is reduced significantly in the land (ocean).

## 4. Summary

In the backdrop of the continued advancement in the coupled climate models, the improvement of seasonal prediction of ISMR is still sub-critical. Previous studies (Kang and Shukla, 2006; Wang et al., 2005) argued that the ability to represent the SST-rainfall relationship is the key to successfully simulating ISM. The cloud-SST relationship is also essential as the formation of clouds is responsible for the rainfall. Therefore, it is timely to evaluate the new generation climate model (CMIP6) due to the scarcity of studies investigating the complex relationship between clouds and SST. The main findings from the study are summarized below:

1. The TCF and convection are essential for the seasonal mean ISMR. The seasonal mean of all the variables (e.g., rainfall, TCF, OLR) is improved in CMIP6-MME and has relatively better agreement with observations/reanalysis over ASM and ISM region.
2. The findings reveal the linkage of observed TCF (and rainfall) over the ISM region with slowly varying forcing (e.g., global SST). The observed/reanalysis teleconnection pattern of TCF-SST is almost similar to that of rainfall-SST. This highlights the interannual variability of clouds, and rainfall is positively coupled.
3. In the long-term period, TCF and SST show solid and positive teleconnection (Fig. 2d) with ET (R ~0.41), NAO (R ~0.51), and AMO (R ~0.41) SST regions in addition to canonical ENSO teleconnection (R ~ -0.39).  This is better captured in CMIP6-MME than CMIP5-MME (Figs. 2, 3). The representation of the global teleconnection pattern has been significantly improved in participating models from CMIP5 to CMIP6 (Figs. S1, S2, 3). The teleconnection with extra-tropics and north Atlantic mode of variability is markedly enhanced in CMIP6-MME than CMIP5-MME. This is invigorating for the seasonal prediction of ISMR.
4. Majority of CMIP6 models are diagonally closer to the top right corner than that of CMIP5 models, indicating the improvement in the bias of both TCF and rainfall (Fig. 4a). The representation of large-scale convection for ISM is also enhanced from CMIP5 to CMIP6 (Fig. 4c). Biases of rainfall, TCF over the ISM region, and OLR over large-scale convection region have been significantly reduced in CMIP6. The dry (wet) bias over land (ocean) in the ASM region has also been improved in CMIP6.

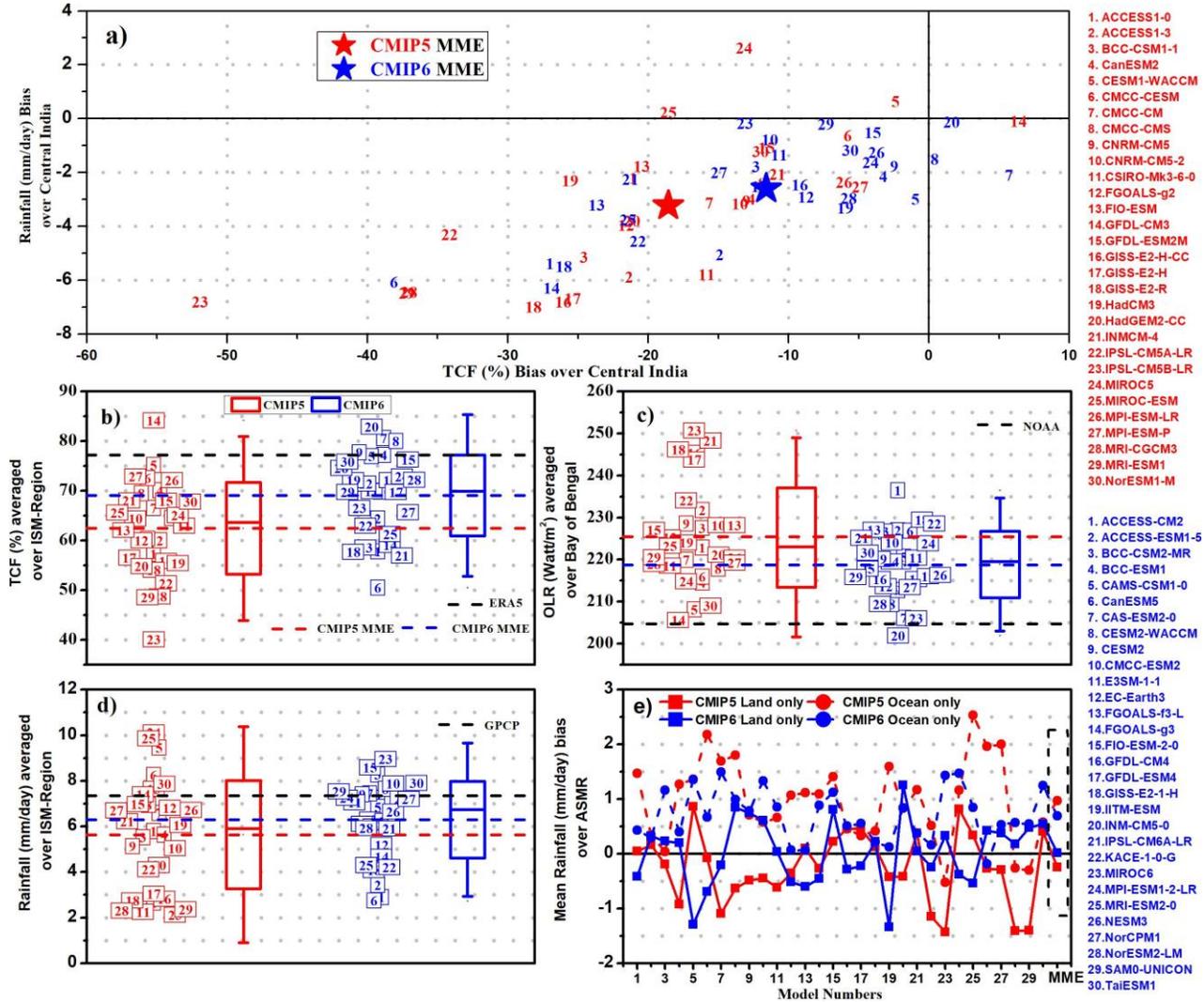

**Figure 4:** *(a) Scatter plot between biases of seasonal mean rainfall and TCF averaged over central India for all thirty models from CMIP5 and CMIP6 and their MMEs. (b) Box and whisker plot of thirty years TCF climatological values averaged over ISM region from all the models. (c) Same as (b), but for OLR averaged over BoB region (d) Same as (b), but for rainfall. In the box and whisker plot here (b-d), the center of each box is the respective MME and is also shown as dashed red(blue) lines for CMIP5(CMIP6) spanned over the whole panel. Observation/reanalysis values are also shown in a black dashed line. The box and whisker boundaries are MME (+/-) 1 and MME (+/-) 2 standard deviations, respectively. The solid line inside each box denotes the statistical median of the ensemble members. Numerical numbers inside tiny rectangles denote the model numbers. (e) Mean rainfall biases over land and ocean separately (over ASM region) are shown from all CMIP5, CMIP6 models and their MMEs.*

Therefore, the improved understanding of the teleconnection of cloud variables with ENSO and other predictors (ET, NAO, and AMO) will help researchers take up the challenges for improving the ISMR skill far ahead using the new generation coupled climate models. This may facilitate reliable seasonal ISM forecasting.

**Acknowledgments**
We thank MoES, the Government of India, and Director IITM for all the support to carry out this work. We acknowledge the climate modeling groups for providing their model output via World Climate Research Programme, and the Earth System Grid Federation (ESGF, https://esgf-node.llnl.gov/), the HadiSST (https://www.metoffice.gov.uk/hadobs/hadisst/), the CALIPSO-GOCCP (https://climatedataguide.ucar.edu/) the KALPANA-1 (https://www.mosdac.gov.in/kalpana-1-introduction), the IMD (https://mausam.imd.gov.in/), the ERA5 (https://cds.climate.copernicus.eu/#!/search?text=ERA5&type=dataset), the GPCP (http://www.esrl.noaa.gov/psd/data/gridded/data.gpcp.html) data sets used here. We also thank the freely available software viz.  The Grid Analysis and Display System (GrADS), NCAR Command Language (NCL), Ferret-NOAA, Climate Data Operators (CDO), and Origin Lab.

**Supplementary Figures:**

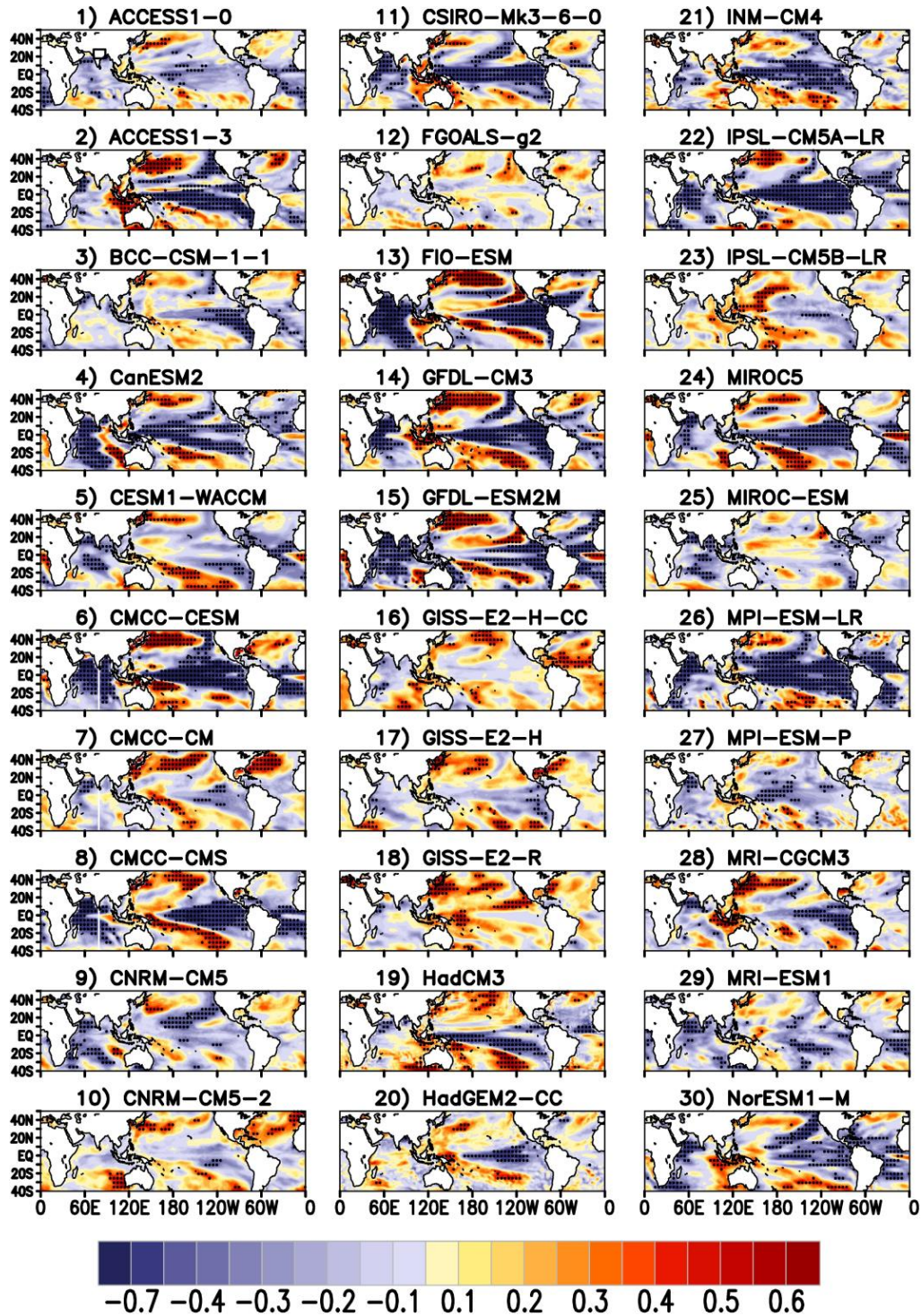

**Figure S1:** *Teleconnection from CMIP5 models*

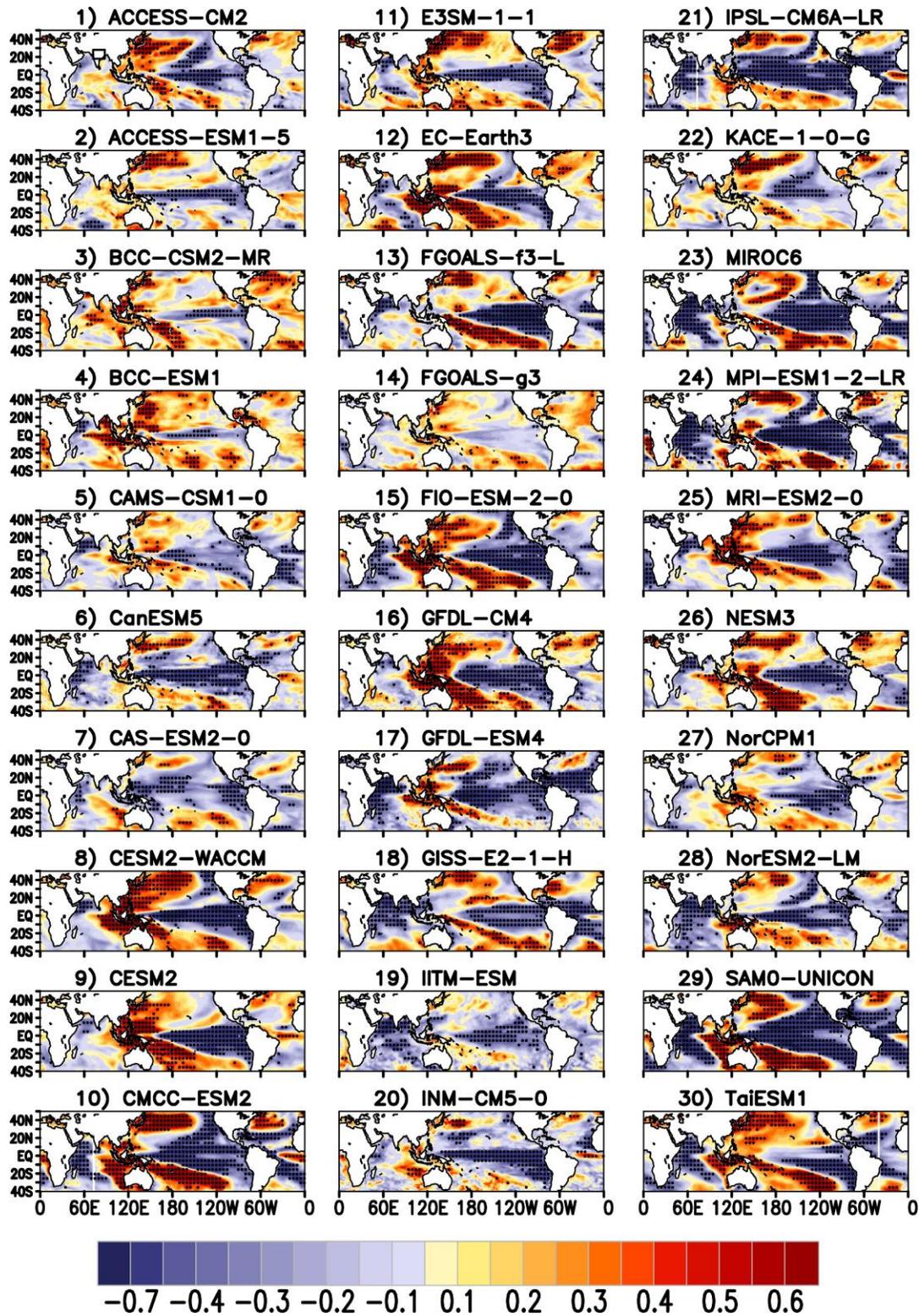

**Figure S2:** *Teleconnection from CMIP6 models*